\documentclass[aps,prd,twocolumn,showpacs,amsmath,amssymb]{revtex4-2}
\usepackage{amsmath}
\usepackage{graphicx}
\usepackage{subfigure}
\usepackage{epstopdf}
\usepackage{color}
\usepackage{multirow}
\usepackage{setspace}
\usepackage{overpic}
\usepackage{amssymb}
\usepackage{adjustbox}

\usepackage{lineno}
\usepackage{bm}
\usepackage{rotating}
\usepackage{makecell}
\usepackage{soul}
\usepackage{float}

\usepackage{hyperref}
\hypersetup{
	colorlinks=true,
	linkcolor=blue,
	filecolor=blue,
	urlcolor=blue,
	citecolor=blue,
}
\let\oldequation\equation
\let\oldendequation\endequation

\renewenvironment{equation}
  {\linenomathNonumbers\oldequation}
  {\oldendequation\endlinenomath}

\begin{document}
\title{\bf \boldmath
Search for the leptonic decay $D^{+}\to e^{+}\nu_{e}$ }

\author{
\small
M.~Ablikim$^{1}$, M.~N.~Achasov$^{4,c}$, P.~Adlarson$^{76}$, O.~Afedulidis$^{3}$, X.~C.~Ai$^{81}$, R.~Aliberti$^{35}$, A.~Amoroso$^{75A,75C}$, Q.~An$^{72,58,a}$, Y.~Bai$^{57}$, O.~Bakina$^{36}$
% ...
\\
\vspace{0.2cm}
(BESIII Collaboration)\\
\vspace{0.2cm} {\it
$^{1}$ Institute of High Energy Physics, Beijing 100049, People's Republic of China\\
% ...
$^{3}$ Bochum Ruhr-University, D-44780 Bochum, Germany\\
$^{4}$ Budker Institute of Nuclear Physics SB RAS (BINP), Novosibirsk 630090, Russia\\
% ...
$^{35}$ Johannes Gutenberg University of Mainz, Johann-Joachim-Becher-Weg 45, D-55099 Mainz, Germany\\
$^{36}$ Joint Institute for Nuclear Research, 141980 Dubna, Moscow region, Russia\\
% ...
$^{57}$ Southeast University, Nanjing 211100, People's Republic of China\\
$^{58}$ State Key Laboratory of Particle Detection and Electronics, Beijing 100049, Hefei 230026, People's Republic of China\\
% ...
$^{72}$ University of Science and Technology of China, Hefei 230026, People's Republic of China\\
% ...
$^{75}$ University of Turin and INFN, (A)University of Turin, I-10125, Turin, Italy; (B)University of Eastern Piedmont, I-15121, Alessandria, Italy; (C)INFN, I-10125, Turin, Italy\\
$^{76}$ Uppsala University, Box 516, SE-75120 Uppsala, Sweden\\
% ...
$^{81}$ Zhengzhou University, Zhengzhou 450001, People's Republic of China\\
\vspace{0.2cm}
$^{a}$ Deceased\\
% ...
$^{c}$ Also at the Novosibirsk State University, Novosibirsk, 630090, Russia\\
% ...
}}


\begin{abstract}
  We  search for the leptonic decay $D^+\to e^+\nu_{e}$ using an $e^+e^-$ collision
  data sample with an integrated luminosity of 20.3~fb$^{-1}$ collected with
  the BESIII detector at the center-of-mass energy of 3.773~GeV,
  No significant signal is observed and an upper limit on the branching fraction
  of $D^+\to e^+\nu_{e}$ is set as $9.7 \times 10^{-7}$, at the 90\%
  confidence level. Our upper limit is an order of magnitude
  smaller than the previous limit for this decay mode.
\end{abstract}

\maketitle

\oddsidemargin  -0.2cm
\evensidemargin -0.2cm

\section{Introduction}
\begin{figure}[htbp]
  \centering
  \includegraphics[width=0.3\textwidth]{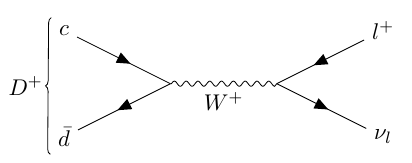}
 \caption{Feynman diagram of $D^{+}\to \ell^+\nu_\ell$.}
  \label{fig:fm}
\end{figure}

Leptonic decays of charmed mesons offer a clean and direct way 
to understand weak decays of the $c$ quark(see e.g.~\cite{Ke:2023qzc} for a recent review). 
The leptonic decays $D^+\to \ell^+\nu_\ell$~($\ell=e$, $\mu$ or $\tau$) 
occur via the annihilation of the $c$ and $\bar{d}$ quarks into
an $\ell^+\nu_\ell$ mediated by a virtual $W^+$ boson, as depicted in
Fig.~\ref{fig:fm}.
The weak and strong interaction effects factorize, leading to a simple expression
for  the partial decay width of $D^+ \to \ell^+\nu_\ell$ at the lowest order
in the Standard Model~(SM).  
It is proportional to the product of the decay constant $f_{D^+}$,
which characterizes the strong-interaction effects between the initial-state quarks,
and the magnitude of the Cabibbo-Kobayashi-Maskawa~(CKM) matrix element $|V_{cd}|$,
representing the $c\to d$ flavor-changing interaction.
In the SM, the decay width can be written as~\cite{decayrate}
\begin{equation}
\Gamma_{D^+\to\ell^+\nu_\ell}=\frac{G_F^2}{8\pi}|V_{cd}|^2
f^2_{D^+} m_\ell^2 m_{D^+} \left (1-\frac{m_\ell^2}{m_{D^+}^2} \right )^2,
\end{equation}
where $G_F$ is the Fermi coupling constant, $m_\ell$ is the lepton mass,
and $m_{D^+}$ is the $D^+$ mass. Thus, the ratio of branching fractions
between different lepton channels depends only on the lepton masses 
and is accurately predicted to be
\begin{equation}
 e^{+}\nu_{e}:\mu^{+}\nu_{\mu}:\tau^{+}\nu_{\tau} = 2.35\times10^{-5}:1:2.67,
 \end{equation}
with negligible uncertainty.  Any observation of violation of this relation 
indicates new physics beyond the SM. 

The $D^+\to e^+\nu_e$ decay, with an expected branching
fraction less than $10^{-8}$, has not yet been observed experimentally. 
The CLEO Collaboration  searched for $D^+\to e^+\nu_e$~\cite{CLEO:2008ffk} and
reported an upper limit of the branching fraction of $8.8\times10^{-6}$ at the 90\%
confidence level using 818~pb$^{-1}$ of the $\psi(3770)$ data. In this paper, we search for $D^+\to e^+\nu_e$ by using 20.3~fb$^{-1}$ of $e^+e^-$
collision data~\cite{BESIII:2024lbn,BESIII:2024lbn2}, approximately 25 times larger than the CLEO measurement, 
collected with the BESIII detector at  the center-of-mass energy of 3.773~GeV.  Charge-conjugate modes are always implied throughout the text.

\section{Description of BEPCII and the BESIII detector}
The BESIII detector\cite{BESIII:2009fln} records symmetric $e^+e^-$ collisions provided by the BEPCII storage ring~\cite{Yu:2016cof} operating
in the center-of-mass energy ($\sqrt{s}$) range from 1.84 to 4.95~GeV, with a peak luminosity of $1.1 \times 10^{33}\;\text{cm}^{-2}\text{s}^{-1}$ 
achieved at $\sqrt{s} = 3.773\;\text{GeV}$. 
BESIII has collected large data samples in this energy region~\cite{BESIII:2020nme,lu2020online,zhang2022suppression}. The cylindrical core of the BESIII detector covers 93\% of the full solid angle and consists of a helium-based
 multilayer drift chamber~(MDC), a plastic scintillator time-of-flight
system~(TOF), and a CsI(Tl) electromagnetic calorimeter~(EMC),
which are all enclosed in a superconducting solenoidal magnet
providing a 1.0 T magnetic field.
The solenoid is supported by an
octagonal flux-return yoke with resistive plate counter muon
identification modules interleaved with steel. 
The charged-particle momentum resolution at $1~{\rm GeV}/c$ is
$0.5\%$, and the ${\rm d}E/{\rm d}x$ resolution is $6\%$ for electrons 
from Bhabha scattering. 
The EMC measures photon energies with a
resolution of $2.5\%$ ($5\%$) at $1$~GeV in the barrel (end-cap)
region. The time resolution in the TOF barrel region is 68~ps, while
that in the end-cap region was 110~ps. The end-cap TOF
system was upgraded in 2015 using multi-gap resistive plate chamber
technology, providing a time resolution of
60~ps~\cite{li2017study,guo2017study,CAO2020163053}.
About 85\% of the data used here benefits from this upgrade.

\section{Monte Carlo simulation}
Monte Carlo (MC) simulated data samples produced with a {\sc
geant4}-based~\cite{geant4} software package, which
includes the geometric description of the BESIII detector and the
detector response, are used to determine detection efficiencies
and to estimate backgrounds. The simulation models the beam
energy spread and initial state radiation (ISR) in the $e^+e^-$
annihilations with the generator {\sc
kkmc}~\cite{kkmc}.
The inclusive MC sample includes the production of $D\bar{D}$
pairs (and treats quantum coherence for the neutral $D$ channels),
non-$D\bar{D}$ decays of the $\psi(3770)$, ISR
production of the $J/\psi$ and $\psi(3686)$ states, and 
continuum processes incorporated in {\sc kkmc}~\cite{kkmc,kkmc2}.
All particle decays are modelled with {\sc
evtgen}~\cite{evtgen,Ping2008zz} using branching fractions 
either taken from the
Particle Data Group (PDG)~\cite{PDG2022} when available,
or otherwise estimated with {\sc lundcharm}~\cite{lundcharm}.
Final state radiation~(FSR)
from charged final state particles is incorporated using  {\sc photos}~\cite{photos2}.
The leptonic decay $D^{+}\to e^{+}\nu_{e}$ is simulated with the {\sc
  SLN} model~\cite{SLN}.  A signal MC sample comprising 5 million
simulated signal events is used to determine the selection efficiencies and model the signal shape.

\section{Analysis Method}
The process $e^{+}e^{-} \to \psi(3770) \to D^{+}D^{-}$, without
accompanying hadrons, allows studies of $D^+$  decays with a double tag technique~\cite{MarkIII-tag, BESIII:2023htx}.
There are two types of samples used in this technique: single tag (ST) and double tag (DT). In the ST sample, a $D^-$ meson is reconstructed via the six hadronic decay modes of
$D^-\to K^{+}\pi^{-}\pi^{-}$, $K^0_{S}\pi^{-}$, $K^{+}\pi^{-}\pi^{-}\pi^{0}$, $K^0_{S}\pi^{-}\pi^{0}$,
$K^0_{S}\pi^{+}\pi^{-}\pi^{-}$, and $K^{+}K^{-}\pi^{-}$.
In the DT sample, both charged $D$ mesons in the event are reconstructed: a ST $D^-$, and 
a signal $D^+ \to e^+ \nu_e$ decay is reconstructed with the remaining tracks.  

The branching fraction of the $D^+\to e^+\nu_{e}$ decay is determined by
\begin{equation}\label{eq:BF}
{\mathcal B}_{D^+\to e^+\nu_{e}} = \frac{N_{\rm DT}}{N^{\rm tot}_{\rm ST} \bar \epsilon_{\rm sig}},
\end{equation}
where $N_{\rm ST}^{\rm tot}$ is the total yield of ST $D^-$ mesons,
$N_{\rm DT}$ is the DT  yield, and
$\bar \epsilon_{\rm sig}$ is the averaged signal efficiency weighted by the ST yields of the $i^{th}$ tag mode in data.  This efficiency is calculated as 
   \begin{equation}
   {
\bar{{\mathcal \epsilon}}_{\rm sig} = \frac{\sum_i (N^i_{\rm ST} \, \epsilon^i_{\rm DT}/\epsilon^i_{\rm ST})}{N^{\rm tot}_{\rm ST}},}
    \end{equation}
where $N^i_{\rm ST}$ is the number of ST $D^-$ mesons for the $i^{th}$ tag mode in data, 
$\epsilon^i_{\rm ST}$ is the efficiency of reconstructing the ST mode $i$, 
and $\epsilon^i_{ \rm DT}$ is the efficiency of finding the tag mode $i$ and the $D^+\to e^+\nu_{e}$ decay simultaneously. 

\section{Particle reconstruction}

All charged tracks detected in the MDC must satisfy $|\cos\theta|<0.93$, where $\theta$ is the polar angle with respect to the $z$-axis, which is the symmetry axis of the MDC. For charged tracks not originating from $K_S^0$ decays, the distance of closest approach to the interaction point (IP) is required to be less than 1\,cm in the transverse plane, and less than 10\,cm along the $z$-axis. Particle identification (PID) for charged tracks combines the ${\rm d}E/{\rm d}x$ measurement in the MDC with the time of flight measurement of the TOF to define the likelihood function $\mathcal{L}(h)~(h=K,\pi, e)$ for each particle ($h$) hypothesis.
Charged kaons and pions are identified by requiring $\mathcal{L}(K)>\mathcal{L}(\pi)$ and $\mathcal{L}(\pi)>\mathcal{L}(K)$, respectively, while positron candidates must satisfy $\mathcal{L}(e)>0.001$ and $\mathcal{L}(e)$/($\mathcal{L}(e)+\mathcal{L}(K)+\mathcal{L}(\pi)$)$>$0.8.
To further reduce mis-identifications between positrons and hadrons, we require $E/p>0.8$, where $E$ is the energy deposit in the EMC from the track and $p$ is its momentum reconstructed in the MDC.
To partially recover the energy loss due to FSR and bremsstrahlung, the four-momenta of photon(s) within $5^\circ$ of the initial positron direction are added to the positron candidate's four-momentum.

The $K_{S}^0$ candidates are reconstructed from pairs of oppositely charged tracks,
each with a distance of closest approach to
the IP less than 20\,cm along the $z$-axis.
The tracks are assigned as $\pi^+\pi^-$ without imposing any PID criteria.
They are constrained to originate from a common vertex and are required to have
an invariant mass within $(0.487,0.511)$~GeV/$c^2$.
The decay length of the $K^0_S$ candidate is required to be greater than
twice the vertex resolution away from the IP.
The quality of both primary and secondary vertex fits is ensured by requiring
$\chi^2$ $<$ 100.  The fitted $K_S^0$ four-vectors are used for later kinematic
calculations.  

The photon candidates are reconstructed from isolated EMC showers. The deposited energy of each shower  in the end-cap region~($0.86 <|\!\cos \theta|< 0.92$)  and in the barrel region~($|\!\cos \theta|< 0.80$) must be greater than 50 MeV or $25$ MeV, respectively. To exclude showers that originate from charged tracks, the angle subtended by the EMC shower and the position of the closest charged track at the EMC must be greater than $10^\circ$  as measured from the IP. The difference between the EMC time and the event start time is required to be within \mbox{[0, 700]~ns} to suppress electronic noise and showers unrelated to the event.

The $\pi^0$ candidates are reconstructed from photon pairs 
with a $\gamma\gamma$ invariant mass within $(0.115,0.150)$\,GeV$/c^{2}$. 
A mass-constrained~(1C) fit is imposed constraining the $\gamma\gamma$ invariant mass to the $\pi^{0}$ nominal mass~\cite{PDG2022} to improve the momentum resolution.
The $\chi^2$ must be less than 50, and the four-momentum of the $\pi^0$ candidate
updated by the fit is used for further analysis. 
 
\begin{figure}
\includegraphics[width=1.01\linewidth]{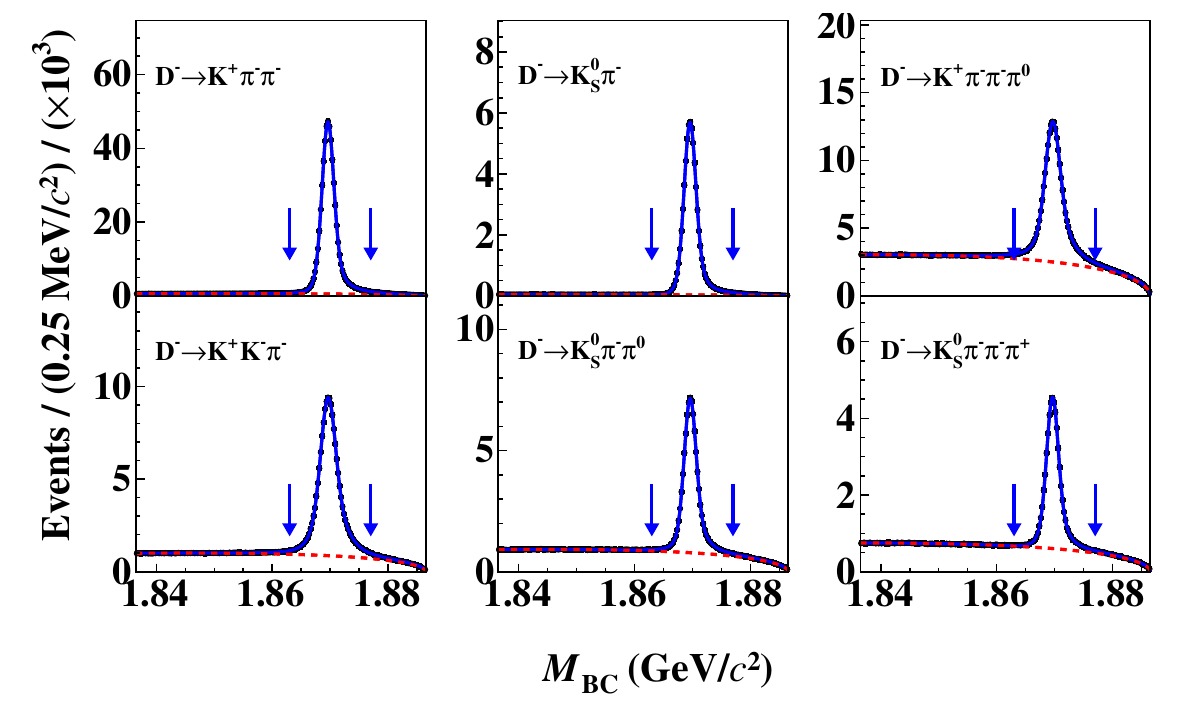}
\caption{Fits to the $M_{\rm BC}$  distributions of the ST $D^{-}$ candidates. The dots with error bars are data, the solid blue lines are  the total fit and the red dashed curves describe the fitted combinatorial background shapes. The pairs of blue arrows indicate the $M_{\rm BC}$ signal window.
}
\label{fig:datafit_Massbc}
\end{figure}

\section{The single-tag selection and yields}
To separate ST $D^-$ mesons from combinatorial backgrounds, we make
use of two kinematic observables, the energy difference  $\Delta E\equiv E_{D^-}-E_{\mathrm{beam}}$ and the beam-constrained mass $M_{\rm BC}\equiv\sqrt{E_{\mathrm{beam}}^{2}/c^{4}-|\vec{p}_{D^-}|^{2}/c^{2}}$, where $E_{\mathrm{beam}}$ is the beam energy, and $E_{D^-}$ and $\vec{p}_{D^-}$ are the energy and momentum of the ST $\bar D$ meson in the $e^+e^-$ center-of-mass frame.
If there is more than one $D^-$ candidate in a given ST mode, the one with the smallest $|\Delta E|$ is kept for further analysis. The $\Delta E$ requirements on the different tag modes are summarized in Table~\ref{ST:realdata}.

For each tag mode, the yield of ST $D^-$ meson is extracted by fitting the corresponding $M_{\rm BC}$ distribution. In the fit, the signal shape is  described by the MC-simulated signal shape convolved with a double-Gaussian function to account for the data-MC resolution difference.
The background shape is described by an ARGUS function~\cite{ALBRECHT1990278}, with the endpoint fixed at $E_{\rm beam}$ = 1.8865~GeV. 
Figure~\ref{fig:datafit_Massbc} shows the fit results for the tag modes in data. The ST efficiencies are obtained by analyzing  the inclusive MC sample.
The candidates with $M_{\rm BC}$ within $(1.863,1.877)$ GeV/$c^2$ are kept for further analysis. 
The ST yields and efficiencies are summarized in Table~\ref{ST:realdata}.

\begin{table}
\renewcommand{\arraystretch}{1.2}
\centering
\caption {The requirements on $\Delta E$, ST $D^-$ yields ($N^i_{\rm ST}$) in data, ST efficiencies ($\epsilon_{\rm ST}^{i}$), and DT efficiencies ($\epsilon^i_{\rm DT}$). The uncertainties on $N^i_{\rm ST}$ are statistical only.}
	\adjustbox{width=0.49\textwidth}{%
\begin{tabular}{lcccc}
\hline
\hline
Tag mode                                                  & $\Delta E$~(MeV)     &  $N^{i}_{\rm ST}~(\times 10^3)$  &  $\epsilon^i_{\rm ST}~(\%)$ &  $\epsilon^i_{\rm DT}~(\%)$      \\\hline
$K^+\pi^-\pi^-$                      &  $[-25,24]$    & $5567.2\pm2.5$      &$50.08$    &$33.92$\\
$K^{0}_{S}\pi^{-}$                   &  $[-25,26]$    & $656.5\pm0.8$      &$51.42$    &$35.00$\\
$K^{+}\pi^{-}\pi^{-}\pi^{0}$         &  $[-57,46]$    & $1740.2\pm1.9$      &$24.53$    &$17.86$\\
$K^{+}K^{-}\pi^{-}$                  &  $[-24,23]$    & $481.4 \pm0.9$      &$41.91$    &$25.46$\\
$K^{0}_{S}\pi^{-}\pi^{0}$            &  $[-62,49]$    & $1442.4\pm1.5$      &$26.45$    &$20.12$\\
$K^{0}_{S}\pi^{-}\pi^{-}\pi^{+}$     &  $[-28,27]$    & $790.2\pm1.1$      &$29.68$    &$20.08$\\
\hline
\hline
          \end{tabular}
          }
\label{ST:realdata}
\end{table}

\section{The double-tag selection and yields}
For the signal side of $D^{+} \to e^+\nu_e$, only the one positron can
be reconstructed.
The neutrino carries away energy and momentum that are not directly detectable,
but may be inferred from four-momentum conservation.  
The recoiling  positron and  $D^{-}$ tag are combined with the known initial-state
four-momentum to achieve this and help select signal events.  A kinematic fit
is performed, constraining the total four-momentum to the
four-momentum of the initial state and constraining
the invariant masses of the $D^-$ tag and the $D^+$ signal to the known $D^\pm$ mass.  
The four-momentum of the missing neutrino is determined by the fit, and 
the $\chi^2$ of this kinematic fit is required to be less than 50. 
To further suppress backgrounds, it is required that there are no extra $\pi^{0}$ 
($N^{\text{extra}}_{\pi^0}=0$) or good tracks ($N^{\text{extra}}_{\text{char}}=0$) 
that are not used in the DT reconstruction. 
The maximum energy of any extra photon ($E_{\rm max,\gamma}^{\rm extra}$)
is also required to be less than 0.2~GeV; this is optimized by maximizing
$\frac{\epsilon}{1.5+\sqrt{B}}$~\cite{fom},  where $\epsilon$ is the signal
efficiency and $B$ is the background yield estimated by the inclusive MC sample.  
The signal yield is determined from a fit to the missing-mass squared, $M_{\rm miss}^2$,
defined as
\begin{equation}
  \begin{aligned}
    M_{\rm miss}^2 &= (E_{\rm beam}-E_{e^+})^2-(-\vec{p}_{D^-}-\vec{p}_{e^+})^2,
  \end{aligned}	
\end{equation}
where $E_{e^+}$ ($\vec{p}_{{e^+}}$) is the energy 
(momentum) of the candidate positron. 

\begin{figure}[!b]
  \centering
  \includegraphics[width=0.45\textwidth]{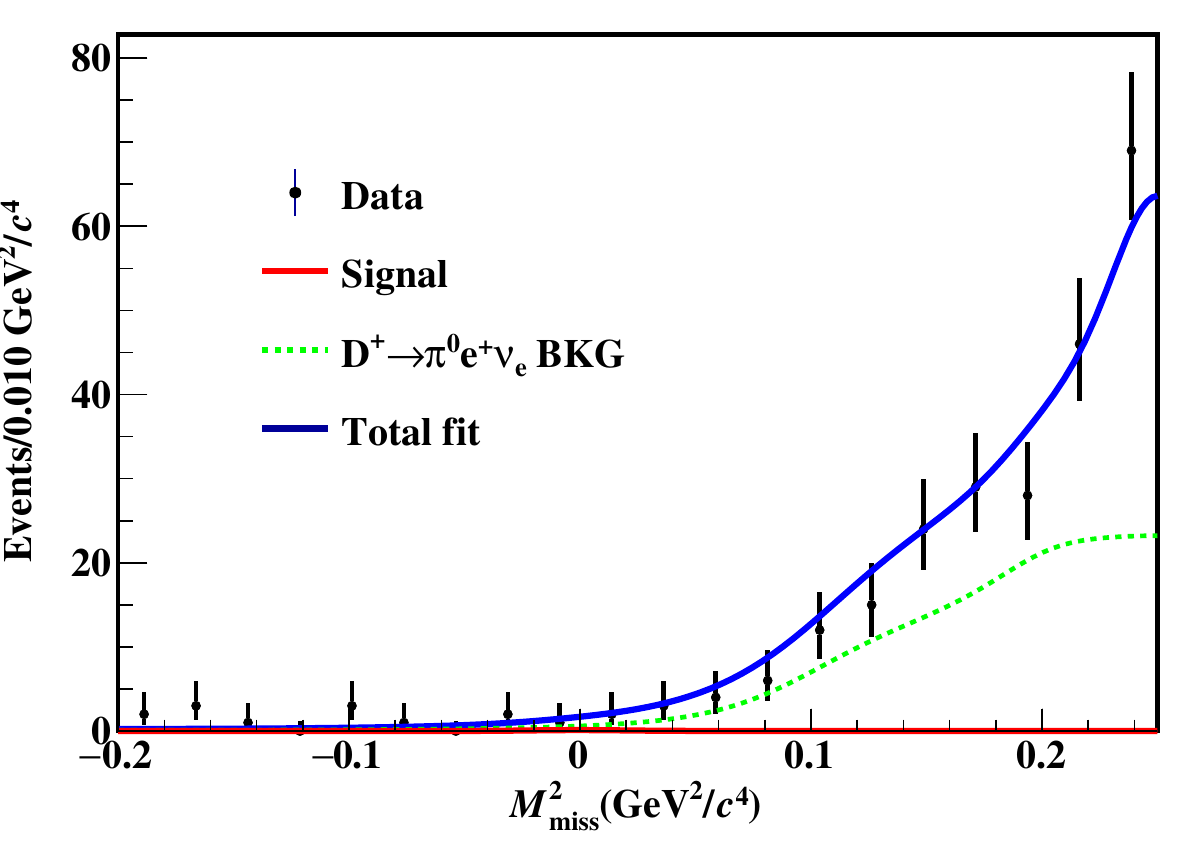}
  \caption{Fit to the $M_{\rm miss}^{2}$ distribution of the accepted candidates for $D^+\to e^+\nu_e$. 
  The dots with error bars are data. The blue solid curve is the fit result.
The red line is the fitted signal shape. }

  \label{fig:crfit}
\end{figure}

We fit the $M_{\rm miss}^2$ distribution in data to obtain the yield of
$D^+\to e^+\nu_e$. The signal shape is derived from the signal MC sample,
and the background shape is derived from the inclusive MC sample, smoothed with the tool
RooKeysPDF~\cite{Roofit}. The decay $D^+ \to \pi^0 e^+ \nu_e$ is the main
background, which is well-modeled in the MC simulation.
The fit result is shown in Fig.~\ref{fig:crfit}; the yield of $D^+\to e^+\nu_e$ is
 $N_{\rm DT}=0.3^{+2.9}_{-3.4}$(stat). 

\section{Systematic uncertainties}

 Most systematic uncertainties related to the efficiency of reconstructing the $D^{-}$ mesons on the tag side are canceled due to the DT method.
The multiplicative systematic uncertainty on the number of single tags, $N_{\rm ST}^{\rm tot}$, is estimated by varying the signal and background shapes, and allowing the parameters of the
   Gaussian to vary in the fit.
It is assigned to be 0.1 $\%$.
The $e^+$ tracking and PID efficiencies  are studied by using a control sample of $e^+ e^- \to \gamma e^+ e^-$. The differences of the efficiencies between data and MC are $1.002 \pm 0.005$ for $e^+$ tracking and $0.972 \pm 0.005$ for PID.
After correcting for the data/MC discrepancy, we assign 0.5\% and 0.5\% as the multiplicative systematic uncertainties
for the $e^{+}$ tracking and PID, respectively.
The efficiency for the combined requirements on $E^{\text{extra}}_{\text{max},\gamma},N^{\text{extra}}_{\text{char}}$ and $N^{\text{extra}}_{\pi^0}$ is studied with a control sample of DT hadronic events where both $D^+$ and $D^-$ decay to one of the six ST hadronic final states.  
The  efficiency difference between data and MC simulation, 1.3\%, is taken as the multiplicative systematic uncertainty.
 We adjust the fit range between (-0.25,0.25) GeV$^2/c^4$ for $M_{\rm miss}^2$, with maximum upper limit and minimum values set as 1.0$\times10^{-6}$ and 9.7$\times10^{-7}$, so we take into account 3\% as additive systematic uncertainty.
All systematic uncertainties are summarized in Table~\ref{tab:sys};
adding them in quadrature results in a total systematic uncertainty of 4.3\%.

\begin{table}[!h]
\begin{center}
\caption{The systematic uncertainties on the branching fraction measurement.}
\begin{tabular}{l|c}
\hline
\hline

{\small Multiplicative Uncertainty } &{\small  Uncertainty~(\%) }\\
\hline
 
{\small  $N_{\rm ST}^{\rm tot}$  }                           & {\small 0.1}\\
 {\small $e^+$ tracking }&{\small  0.5}\\
 {\small $e^+$ PID  }                       & {\small 0.5}\\
{\small $E^{\text{extra}}_{\text{max},\gamma}, N^{\text{extra}}_{\text{char}}, N^{\text{extra}}_{\pi^0}$
 }                           & {\small 3.0}\\
\hline
{\small Additive  Uncertainty    }                               &  {\small Uncertainty~(\%)}\\\hline
{\small $M_{\rm miss}^2$  fit region }&{\small  3.0}\\
\hline

{\small Total   }                               &  {\small 4.3}\\

\hline
\hline
\end{tabular}
\label{tab:sys}
\end{center}
\end{table}
  
\begin{figure}[!h]
\centering
  \includegraphics[width=0.45\textwidth]{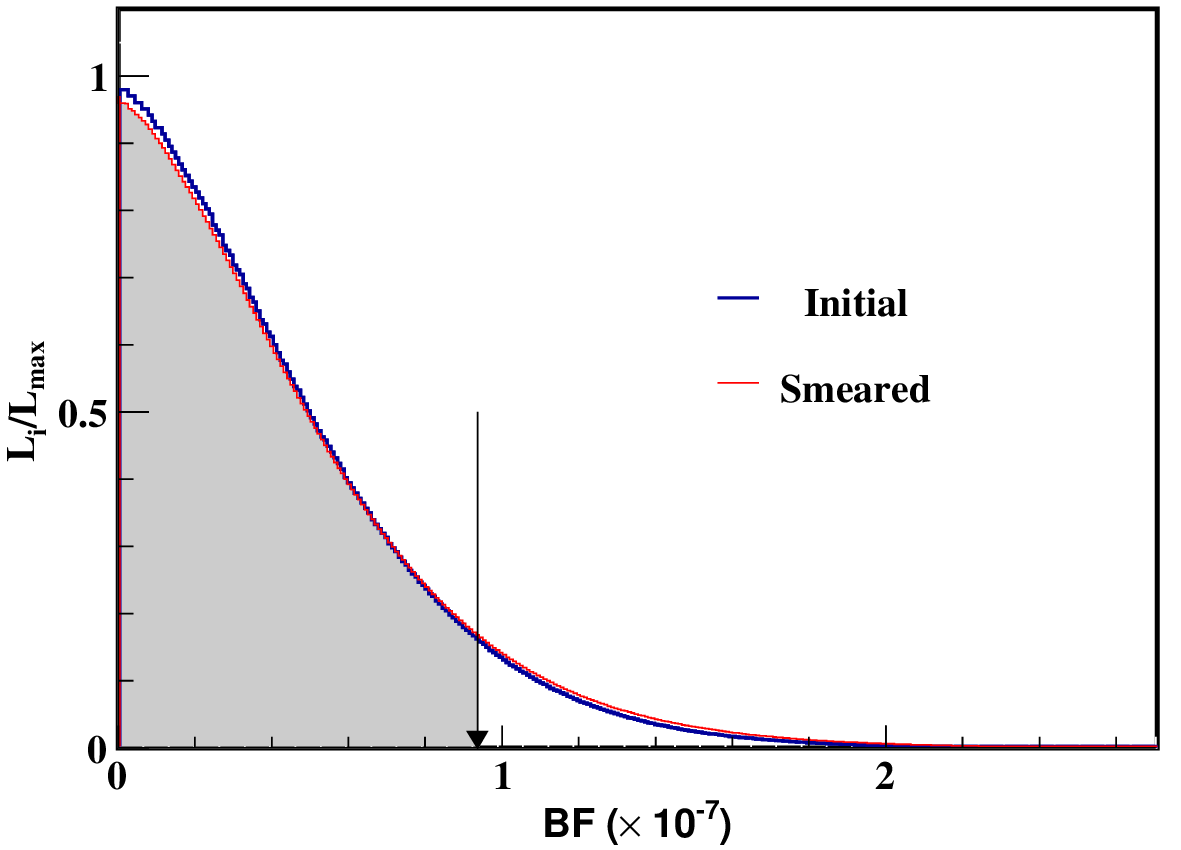}
  \caption{Distribution of likelihood versus the branching fraction of $D^{+} \to e^+ \nu_{e}$.
    The likelihood in each  bin is denoted as $L$ and the maximum of the likelihood is $L_{\text{max}}$.
  The results obtained with and without incorporating the systematic
  uncertainties are shown as the red and blue curves, respectively. The black arrow
  shows the upper limit corresponding to the 90\% confidence level.
  }
  \label{fig:up}
\end{figure}
\section{Upper limit of the branching fraction}
Since no significant signal is found, an upper limit on the branching fraction of $D^+\to e^+\nu_e$ is  estimated using the Bayesian approach. 
The sources of systematic uncertainties on the upper limit measurements are
classified into two types: additive and
multiplicative~($\sigma_s$).
To incorporate the multiplicative systematic uncertainty in
the calculation of the upper limit,
the likelihood distribution
is smeared by a Gaussian function with a mean of
zero and a width equal to $\sigma_{s}$ as described below~\cite{Stenson},
\begin{equation}\label{eqBr0_BF2}
  L(\mathcal{B}) \propto {\displaystyle{{\int}^{1}_{0}} \textstyle{ L \left( \frac{\epsilon_S}{\epsilon_{\hat{S}}}\mathcal{B} \right) e^{ \left[ - \frac{\left(\epsilon_S - \epsilon_{\hat{S}}\right)^2}{2 {\sigma}^2_{S}} \right]} d S }},
\end{equation}
where we associate $\epsilon_{\hat{S}}$ with the nominal efficiency, $\epsilon_{S}$ with  the expected efficiency and $L(\mathcal{B})$ is the likelihood distribution obtained from a fit to the likelihood of $\mathcal{B}$~(branching fraction) and parameterized as a Gaussian. The only significant additive uncertainty comes from the normalization of the
$D^+ \to \pi^0 e^+ \nu_e$ branching fraction.   
We repeat the maximum-likelihood fit, varying this BF by the PDG uncertainty
and choose the most conservative upper limit among these results. 
The distribution of the likelihood versus the assumed branching fraction is shown in Fig.~\ref{fig:up}. Finally, the upper limit on the branching fraction of $D^+\to e^+\nu_e$ at the 90\% confidence level is set at 9.7$\times 10^{-7}$.

\section{Summary}
In summary, by analyzing 20.3~fb$^{-1}$ of $e^+e^-$ collision data
collected at $\sqrt s=3.773$~GeV with the BESIII detector,
 we search for
the leptonic decay $D^+\to e^+\nu_{e}$. No significant signal is  observed and an upper limit on the branching fraction of $D^+\to e^+\nu_e$ is set at $9.7\times10^{-7}$
at the 90\% confidence level. The sensitivity is improved by an order of magnitude compared to the CLEO measurement.

\section{Acknowledgement}

The BESIII Collaboration thanks the staff of BEPCII and the IHEP computing center for their strong support. This work is supported in part by National Key R\&D Program of China under Contracts Nos. 2020YFA0406400, 2020YFA0406300, 2023YFA1606000; National Natural Science Foundation of China (NSFC) under Contracts Nos. 11635010, 11735014, 11935015, 11935016, 11875054, 11935018, 12025502, 12035009, 12035013, 12061131003, 12192260, 12192261, 12192262, 12192263, 12192264, 12192265, 12221005, 12225509, 12235017, 12361141819; the Chinese Academy of Sciences (CAS) Large-Scale Scientific Facility Program; the CAS Center for Excellence in Particle Physics (CCEPP); Joint Large-Scale Scientific Facility Funds of the NSFC and CAS under Contract Nos. U2032104, U1832207; 100 Talents Program of CAS; The Excellent Youth Foundation of Henan Scientific Commitee under Contract No. 242300421044; The Institute of Nuclear and Particle Physics (INPAC) and Shanghai Key Laboratory for Particle Physics and Cosmology; German Research Foundation DFG under Contract No. FOR5327; Istituto Nazionale di Fisica Nucleare, Italy; Knut and Alice Wallenberg Foundation under Contracts Nos. 2021.0174, 2021.0299; Ministry of Development of Turkey under Contract No. DPT2006K-120470; National Research Foundation of Korea under Contract No. NRF-2022R1A2C1092335; National Science and Technology fund of Mongolia; National Science Research and Innovation Fund (NSRF) via the Program Management Unit for Human Resources \& Institutional Development, Research and Innovation of Thailand under Contracts Nos. B16F640076, B50G670107; Polish National Science Centre under Contract No. 2019/35/O/ST2/02907; Swedish Research Council under Contract No. 2019.04595; The Swedish Foundation for International Cooperation in Research and Higher Education under Contract No. CH2018-7756; U. S. Department of Energy under Contract No. DE-FG02-05ER41374

\bibliography{bibliography}
\end{document}